\documentclass[seceqnum,aps,nofootinbib]{revtex4}

\usepackage{amsfonts}
\usepackage{graphicx}

\newcommand{\X}{X}
\newcommand{\B}{\beta}
\newcommand{\A}{A}
\newcommand{\eqref}[1]{(\ref{#1})}
\newcommand{\BETA}{\epsilon ( \hat{R} _{, \tau} ^{2} + A ^{2} ) ^{1/2}}

\newcommand{\logLr}{\log\left({(\hat R_{,\tau}^2+A^2)^{1/2}+\epsilon \hat
                    R_{,\tau}\over A}\right)}
\newcommand{\logEr}{\log\left({(A^2-\hat R_{,\bar \tau}^2)^{1/2}+i \epsilon 
\hat
                    R_{,\bar \tau}\over A}\right)}

\newcommand{\AS}{A_{\mathrm S}}
\newcommand{\AM}{A_{\mathrm M}}
\newcommand{\tS}{t_{\mathrm S}}
\newcommand{\tM}{t_{\mathrm M}}
\newcommand{\tSi}{t_{\mathrm S}^{\mathrm i}}
\newcommand{\tSf}{t_{\mathrm S}^{\mathrm f}}
\newcommand{\Nt}{N^t}
\newcommand{\Nr}{N^r}
\newcommand{\Ht}{{\cal H}_t}
\newcommand{\Hr}{{\cal H}_r}

\newcommand{\Rg}{R_{\mathrm{g}}}
\newcommand{\rg}{r_{\mathrm{g}}}

\begin{document}
\title{Tunnelling with wormhole creation}
\author{Stefano Ansoldi$^{1,2}$ and Takahiro Tanaka$^{3,4}$}
\affiliation{\,\\ \,\\
$^1$INFN, Sezione di Trieste\\
$^2$Universit\`{a} degli Studi di Udine, via delle Scienze 206, I-33100 Udine (UD), Italy\\
$^3$Department of Physics, Kyoto University, Kyoto, 606-8502, Japan\\
$^4$Yukawa Institute for Theoretical Physics, Kyoto University, Kyoto, 606-8502, Japan\\
}
\date{\today}

\begin{abstract}
The description of quantum tunnelling in the presence of gravity shows subtleties
in some cases. Here we discuss wormhole production in the context of the spherically
symmetric thin-shell approximation. By presenting a fully consistent treatment based
on canonical quantization, we solve a controversy present in literature. 
\end{abstract}

\maketitle
\section{Introduction}

Quantum tunnelling plays various roles in cosmology. 
For instance, false vacuum decay through quantum tunnelling~\cite{Coleman,CC,CdL}  
is an important process for the universe to visit 
many vacua in the string landscape~\cite{GSVW,Carroll,BSV}. 
Also, the possibility of creation of an open universe through 
false vacuum decay has been extensively discussed~\cite{Gott,Yamamoto,Bucher1,Bucher2}. 
Taking properly into account the effect of gravity can be quite non-trivial.
Although in some cases the effect of gravity is secondary, there are in fact
several cases in which gravity plays a crucial role, such as the upward quantum 
tunnelling from a lower to a higher energy vacuum~\cite{LW,Jaume}. 

Even when the effect of gravity is secondary, including gravity 
can make the treatment highly non-trivial. One example is 
the subtle issue raised by Lavrelashvili, Rubakov and Tinyakov~\cite{Lav}
that fluctuations around bubble nucleation might cause an instability, which
leads to explosive particle production. One prescription to cure this pathology
was proposed in Ref.~\cite{Tan92,Tan94}, in which it is shown that one can eliminate
the instability, at least apparently, by an appropriate choice of the gauge. 

Quantum tunnelling in connection with gravity has been discussed
also in other contexts. One of them is wormhole
formation~\cite{Kyo1,Kyo2,Kyo3,Kyo4,Kyo5,Kyo6,Blau,Berezin1,Berezin2}, 
which is the main subject of this paper.
Wormhole formation is a signature of what in the literature is also
referred to as baby/child universe creation~\cite{guend}.
Spherical thin shells with various equations
of state have been studied, as models of matter fields able to describe this process.
Even in the simple case of a pure tension shell, the quantum mechanical formation of
a wormhole seems possible. However, some inconsistencies between different prescriptions
seem to exist in literature~\cite{Stefano}. 
In this paper we will clarify that the origin of these apparent discrepancies 
is tightly related to the use of the time coordinate in the static chart. 
We then propose a plausible prescription based on a smooth time-slicing to tackle
the problem.

This paper is organized as follows. In Sec.~\ref{sec:conventional} we briefly review 
the derivation of the standard result for the tunnelling amplitude
based on the direct evaluation of the action, when the time slice of
the static chart is used.
In Sec.~\ref{sec:wormhole} we discuss the problem that arises when we try to apply the 
conventional formula to situations characterized by wormhole production. 
To overcome some difficulties that appear in this last case, in Sec.~\ref{sec:CASTS}
we then study the same problem using a canonical approach with smooth time slice:
this allows us  to derive the formula for the tunnelling rate without any ambiguity. 
In Sec.~\ref{sec:CDE} we finally show how the same formula can be reproduced by 
the direct evaluation of the action if we carefully take the smooth 
time slice. 
Sec.~\ref{sec:sumcon} is devoted to summary and discussion: we also elaborate on a
remaining, more subtle, issue. 

\section{\label{sec:conventional}Conventional approach}

In this paper we consider the simplest spherically symmetric domain wall model, 
whose Lagrangian is given by 
\begin{equation}
 S={1\over 16\pi G}\int d^4 x\sqrt{-g}\,{\cal R} - 
\int d\tau\, m(\hat R)~, 
\end{equation}
where ${\cal R}$ is the scalar curvature and $m(\hat R)$ 
is the radius dependent mass of the wall, {\it e.g.}, 
$m(\hat R)=$constant for a dust domain wall while
$m(\hat R)=4\pi \sigma \hat R^2$ for a wall consisting of pure tension $\sigma$; 
moreover $\tau$ is the proper time along the wall, while 
$\hat R$ denotes the circumferential radius of the wall. 
In general, quantities marked with a ``$\:\hat{~}\:$''
are considered to be evaluated at the position of the wall, {\it e.g.}, 
$ \hat{B} = \hat{B} ( t ) = B ( t , \hat{r} ( t ) )$ 
when $B$ is a function of $t$ and $r$, and $r = \hat{r} ( t )$ is one possible
parametrization of the wall trajectory.
Depending on the model parameters, the wall motion can have some classically 
forbidden region for a range of the radius. We are interested in discussing
the quantum tunnelling of the wall when it reaches a turning point, \textit{i.e.}
a boundary of the classically forbidden region, by explicitly taking into account
gravity.  

In this section we derive a conventional but incorrect formula for the
tunnelling rate of the wall. 
Although we mostly follow Ref.~\cite{Farhi}, we are \emph{not} claiming 
that the result obtained in this reference is wrong. Indeed, our emphasis
is about the fact that in this reference the authors clearly identified a
discrepancy between the direct evaluation of the action that they propose
and a naive canonical approach. 
Moreover, it was clearly emphasized in Ref.~\cite{Farhi} that the proposed direct
approach guarantees, instead, a continuous variation of the action as the parameters 
(the Schwarzschild mass, the Schwarzschild Sitter cosmological constant, the wall
surface tension, in their model) are changed: on the contrary, the
conventional canonical approach
does not guarantee continuity of the action as a function of the parameters. 
At the same time the direct calculation of the action discloses the difficulties
in the identification of the Euclidean manifold interpolating between the
before- and after-tunnelling classical solutions in a consistent way: indeed,
Farhi et al. associate what they call a \emph{pseudo-manifold} to the
instanton solution. The direct approach defines the pseudo-manifold
by weighting different volumes of the instanton along the classically
forbidden trajectory by an integer number that counts how many times (and
in which direction) the Euclidean volume is swept by the time slice.
We will later show that the canonical approach, in full generality, can
reproduce the same value for the tunnelling action given in the approach
proposed by Farhi et al..

The direct evaluation of the action is possible because the solution is simply
given by a junction of two spacetimes. Here, for simplicity, we assume that the
inside and outside of the bubble are both empty, so that the inside can be taken
as a piece of Minkowski spacetime and the outside as a piece of Schwarzschild
spacetime. (In Ref.~\cite{Farhi} the inside was equipped with vacuum energy density,
\emph{i.e.}, a cosmological constant, but this does not change the treatment in any substantial
way.) The method proposed by Farhi et al. was developed in coordinates
adapted to the static and spherically symmetric nature of the spacetimes
participating in the junction. With this, we mean that the Lagrangian was
preferably considered in connection with the coordinate times in 
the static chart in both spacetime regions, which we denote by $\tS$ and $\tM$
in the simplified case that we are considering here. 
However, most of the calculations were performed using the proper time of an observer
sitting on the junction, and therefore the result can be easily extended to a
coordinate-independent expression, as we shall see in Sec.~\ref{sec:CDE}. 

The contributions to the action can be summarized as follows.
\begin{enumerate}
  \item A matter term coming from the shell, $I ^{\mathrm{wall}} _{\mathrm{matter}}$:
    this is nothing but the contribution from
    the stress-energy tensor localized on the bubble surface.
  \item A gravity term coming from the bubble wall, $I ^{\mathrm{wall}} _{\mathrm{gravity}}$:
    this is, basically, the well-known \emph{extrinsic-curvature-trace-jump} term.
  \item The bulk contributions vanish for classical solutions since there is no matter
    in the bulk. 
  \item Surface terms: although the appearance of surface terms is conceptually clear,
    the treatment of these terms may be non-trivial. As clearly discussed in
    Ref.~\cite{Farhi} several contributions arise.
    \begin{enumerate}
      \item In their treatment, a crucial contribution, 
        $I ^{\mathrm{wall}} _{\mathrm{surface}}$, comes from the bubble wall positions, where
        the normal to the constant time surface is discontinuous. However, this contribution
	does not appear if we adopt a smooth foliation of time across the 
        wall. In Sec.~\ref{sec:CASTS} we take this latter picture. 
      \item Another contribution comes from a surface at a large constant circumferential
        radius in the outside spacetime, $I _{\mathrm{surface}} ^{R _{\mathrm{BIG}}}$:
        this cut-off radius allows us to work with a (spatially) bounded volume,
        and the large radius limit has to be taken in the end.
        This limit naturally brings in divergences, which can be usually dealt with,
        \emph{e.g.}, by the Gibbons-Hawking prescription.
        The final regularized result is called
        $I _{\mathrm{net}} ^{R _{\mathrm{BIG}}}$ below.
    \end{enumerate}
\end{enumerate}
With the notation used above, and by setting (because of the square, the notation below
differs from the one used in Ref.~\cite{Farhi})
\begin{equation}
  A _{\mathrm{M}} ^{2} = 1~ 
  , \quad
  A _{\mathrm{S}} ^{2} = 1 - \frac{2G M}{R}~
  ,
\label{eq:metfun}
\end{equation}
the above terms are~\cite{Farhi} 
\begin{eqnarray}
  I _{\mathrm{matter}} ^{\mathrm{wall}}
  & = &
  -
  \int _{\tau ^{\mathrm{i}}} ^{\tau ^{\mathrm{f}}}
    m ( \hat{R} ) d \tau~,
  \\
  I _{\mathrm{gravity}} ^{\mathrm{wall}}
  & = &
  \int _{\tau ^{\mathrm{i}}} ^{\tau ^{\mathrm{f}}} d \tau
    \left\{
      \frac{1}{2 G}
      \left[
        2 \hat{R} \BETA
        +
        \frac{\hat{R} ^{2}}{\BETA}
        \left(
          \hat{R} _{, \tau \tau}
          +
          \frac{1}{2} (A ^{2}) _{,R}
        \right)
      \right]
    \right\}~,
\label{gravityterm}
  \\
  I _{\mathrm{surface}} ^{\mathrm{wall}}
  & = &
  -
  \frac{1}{2 G}
  \int _{\tau ^{\mathrm{i}}} ^{\tau ^{\mathrm{f}}} d \tau
    \frac{d}{d \tau}
    \left[
      \hat{R} ^{2}
    \logLr
    \right]
    \qquad \qquad \mathrm{(please,\ check\ note\ \footnotemark{})}
  \nonumber \\
  & = &
  -
  \frac{1}{2 G}
  \int _{\tau ^{\mathrm{i}}} ^{\tau ^{\mathrm{f}}} d \tau
    \left[
      2 \hat{R} \hat{R} _{, \tau}
      \logLr +
    \right .
  \nonumber \\
  & & \qquad \qquad
    \left.
      +
      \frac{\hat{R} ^{2}}{\BETA}
      \left( \hat{R} _{, \tau \tau} + \frac{( A ^{2} ) _{, R}}{2} \right)
      -
      \frac{\hat{R} ^{2} \BETA}{2 A ^{2}} ( A ^{2} ) _{, R}
    \right]~,
    \label{eq:actwallsurfin}
\\
  I _{\mathrm{surface}} ^{R _{\mathrm{BIG}}}
  & = &
  \left( \frac{R _{\mathrm{BIG}}}{G} - \frac{3 M _{\infty}}{2} \right)
  \left( t _{\mathrm{S}} ^{\mathrm{f}} - t _{\mathrm{S}} ^{\mathrm{i}} \right)~,
  \\
  I _{\mathrm{net}} ^{R _{\mathrm{BIG}}}
  & = &
  I _{\mathrm{surface}} ^{R _{\mathrm{BIG}}}
  -
  (I _{\mathrm{surface}}) _{0}
  =
  - \frac{M _{\infty}}{2}
  \left( t _{\mathrm{S}} ^{\mathrm{f}} - t _{\mathrm{S}} ^{\mathrm{i}} \right)
  +
  \mathrm{O} \left( \frac{1}{R _{\mathrm{BIG}}} \right)~,
\end{eqnarray}
where square brackets represent the jump of the bracketed quantities across
the shell, {\it i.e.}, 
\begin{equation}
  \left [ \hat{B} \right ]
  =
  \lim _{\delta \to 0^{+}}
    \left( 
      \hat{B} ( \hat{r} - \delta )
      -
      \hat{B} ( \hat{r} + \delta )
    \right)~
    .
\end{equation}
\footnotetext{The expression given in Ref.~\cite{Farhi} looks slightly different,
but it is equivalent to this one as long as we require
that $I^{\mathrm{wall}}_{\mathrm{surface}}$ is always 
real valued. As we shall explain later (see Eq.~\eqref{signflippoint}), the sign flip
of $\epsilon$ is only important in the Euclidean regime. As the 
argument of the logarithm has a jump there, we may have 
to add one more term proportional to a $\delta$ function at the 
sign flipping point to the right hand side of Eq.~\eqref{eq:actwallsurfin}. However, the 
crucial point is that analyticity of $I^{\mathrm{wall}}_{\mathrm{surface}}$ is 
broken at the sign flipping point. Therefore, it is difficult to find 
a consistent meaning for the analytic continuation of this expression 
to the Euclidean region.}
Square brackets will be nowhere used with a different meaning.
Moreover, the signs
\begin{equation}
 \epsilon_\pm = {\rm sign}\left(\AM^2-\AS^2\mp {G^2m^2\over \hat R^2}\right), 
\end{equation}
are unambiguously determined by the consistency with the junction condition~\cite{Israel}
\begin{eqnarray}
{Gm \over \hat R} 
  =
  \left[\epsilon
  \left(
    \hat{R} _{, \tau} ^{2} + {\A} ^{2}
  \right) ^{1/2}\right]~. 
\label{junctionc}
\end{eqnarray}

Noticing that 
\begin{equation}
{d\tS\over d\tau}={\epsilon   \left(
    \hat{R} _{, \tau} ^{2} + {A}_{\mathrm{S}}^{2}
  \right) ^{1/2}\over \AS^2}~,  
\label{dtdtau}
\end{equation}
all the above contributions can be combined to give the Lagrangian
\begin{equation}
  L
  =
  \frac{1}{G}
  {d\tau \over d\tS}
  \left(
    \left\{
    \hat{R}
    \left [ \BETA \right ]
    -
    m ( \hat{R} )
  \right\}
  -
  \hat{R} \hat{R} _{, \tau}
  \left[
    \logLr
  \right]\right)
  -
  M
  ~.
\label{eq:res}
\end{equation}
Finally, adding a constant $M$ to the Lagrangian so that the
Lagrangian vanishes at the turning point, $\hat R_{, \tau} = 0$,  
we can evaluate $L$ on a classical solution to obtain
\begin{equation}
  \left . L \right | _{\mathrm{solution}} 
  =
  -
  \frac{\hat{R} \hat{R} _{, \tS}}{G}
  \left[
    \logLr
  \right]
  .
\label{eq:patintEucmom}
\end{equation}
Here $\hat{R} _{, \tau}$ is to be replaced with its classical solution, which is 
obtained from the junction condition~\eqref{junctionc} as
\begin{equation}
 \hat R_{,\tau}^2
 ={G^2m^2\over 4\hat R^2}
   \left\{1-{(\AS + \AM)^2 \hat R^2\over G^2m^2} \right\}
    \left\{1-{(\AS - \AM)^2 \hat R^2\over G^2m^2} \right\}. 
\label{solRdot}
\end{equation}

As explicitly seen above, the action could in general contain second derivative terms. 
These second derivatives are removed by the ``careful'' inclusion of the boundary
term, $I _{\mathrm{surface}} ^{\mathrm{wall}}$.
From Eq.~(\ref{eq:patintEucmom}), we identify the effective momentum 
conjugate to $\hat R$ as 
\begin{equation}
 P_{\mathrm{eff}} :=  -
  \frac{\hat{R} }{G}
  \left[
   \logLr
  \right]~.
\label{eq:effmomlor}
\end{equation}
After Wick rotation to Euclidean time, 
$\bar\tau=i\tau$, the Euclidean momentum, 
$\bar{P}_{\mathrm{eff}} =-i P_{\mathrm{eff}}$
and Eq.~\eqref{solRdot} become
\begin{equation}
 \bar{P}_{\mathrm{eff}} 
  =  i
  \frac{\hat{R} }{G}
  \left[
   \logEr
  \right]~,
\label{eq:effmomEuc}
\end{equation}
and
\begin{equation}
 \hat R_{,\bar\tau}^2
 ={G^2m^2\over 4\hat R^2}
   \left\{{(\AS + \AM)^2 \hat R^2\over G^2m^2}-1 \right\}
    \left\{1-{(\AS - \AM)^2 \hat R^2\over G^2m^2} \right\}~, 
\end{equation}
respectively. We indicates quantities after Wick rotation with
``$\:\bar{~}\:$'', if they are different from the Lorentzian ones.  
We also note that $\bar P_{\mathrm{eff}}$ is real, since the modulus
of the argument inside the logarithm is unity. 
Then, the tunnelling action may be evaluated as 
\begin{equation}
 \bar I_{(\tS)}=\int d\bar \tS \hat R_{,\tS} \bar P_{\mathrm{eff}}~, 
\label{Iconventional}
\end{equation}
to provide the tunnelling rate $\sim \exp(-2\bar I_{(\tS)})$. 

\section{\label{sec:wormhole}Wormhole production}

\begin{figure}
\begin{center}
\includegraphics[width=8cm]{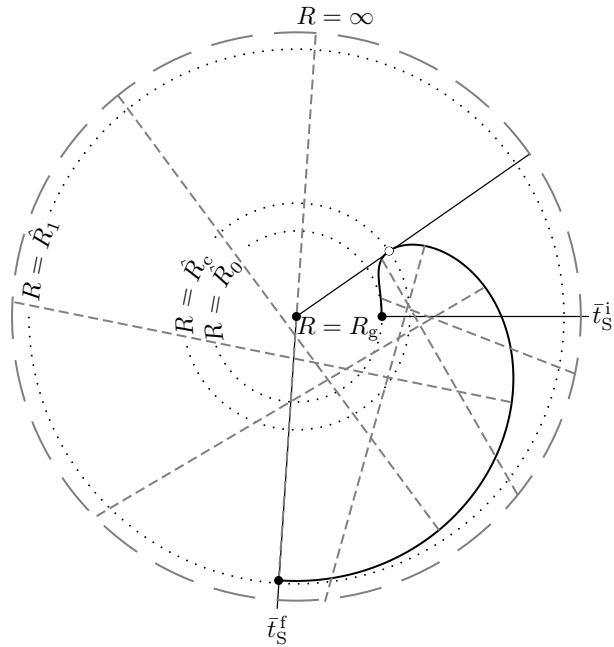}
\end{center}
\caption{\label{fig1}A schematic diagram of 
 the Euclidean Schwarzschild spacetime. The center and the boundary 
  of the circle correspond to $R=\Rg$
 and $R=\infty$, respectively. Dotted circles show the surface $R = \hat{R} _{\mathrm{c}}$
 and those corresponding to the radii of the turning points, $R = \hat{R} _{0,1}$.
 The angle represents the direction of the 
 time coordinate of the static chart, $\bar \tS$. The solid curve represents the
 trajectory of the domain wall, for the Minkowski--Schwarzschild case with
 $M = 1$, $\tilde{\sigma} = 0.25$. Surfaces with $\bar \tS=$constant are
 shown by solid lines. The foliation by these surfaces starts with
 $\bar\tS=\bar\tSi$ 
  and the angle increases at the beginning. After reaching 
 the maximum, the angle starts to decrease to reach
 $\bar\tS=\bar\tSf$.  
 The foliation corresponding to a smooth
 time slicing is presented by dashed lines.}
\end{figure}
The framework discussed in the preceding section 
is generically applicable to the tunnelling problem.  
However, analytic continuation 
brings up situations that are technically
and conceptually more involved. 
To see this, first we notice 
that $\epsilon=\pm 1$ flips sign when 
\begin{equation}
 A^2_\pm+\hat R_{,\tau}^2
 ={G^2m^2\over 4\hat R^2}\left(
  1 \pm {(\AS^2-\AM^2)\hat R^2\over G^2m^2}\right)^2
 ={G^2m^2\over 4\hat R^2}\left(1\mp {R_{\mathrm{g}} \hat R\over G^2m^2}\right)^2
\label{signflippoint}
\end{equation}
vanishes, where $\Rg :=2GM$. 
We denote the value of $\hat R$ at the sign changing point by $\hat R_{\mathrm{c}}$. 
In the Lorentzian regime, the sign flip of $\epsilon$ does not occur in regions
outside horizons: it can happen behind horizons, but in these cases no pathology
arises~\cite{Stefano}. In any case, in this work, because of our definitions
\eqref{eq:metfun}, we are implicitly excluding regions behind horizons. This is
certainly non restrictive for our current purpose, because it is possible to
prove that tunnelling \emph{must always begin and end} in regions that are \emph{not}
behind horizons, and it is always true that $P_{\mathrm{eff}}$ is continuous
during the time evolution.
However, in the Euclidean regime, not only the sign flip can happen, but 
the argument of the logarithm (and hence the logarithm itself) in $P_{\mathrm{eff}}$ also
has a  jump at the point where the sign of $\epsilon$ flips: this can not be
avoided if one consistently requires that the effective momentum should vanish 
at both turning points. (In fact, the discontinuity cannot be avoided if we
require that $P_{\mathrm{eff}}$ analytically continued back to the Lorentzian
regime is real, both, before and after the tunnelling.) This happens because
the expression for $P_{\mathrm{eff}}$ is essentially non-analytic. For this reason,
it is hard to justify the use of analytic continuation of an action that contains
$P_{\mathrm{eff}}$. 

In the present case, from Eq.~\eqref{signflippoint}, we find that the sign 
flip can happen for $\epsilon_+$ only.
 
From the analytic continuation of Eq.~\eqref{dtdtau},
\begin{equation}
{d\bar \tS\over d\bar\tau}={\epsilon   \left(
{A}_{\mathrm{S}}^{2} - \hat{R} _{, \bar\tau} ^{2}
  \right)^{\! 1/ 2}\!\! \over \AS^2}~,  
\end{equation}
we find that $d\tilde \tS/d\tau$ also vanishes at the sign flip point.  
This means that the trajectory of the wall becomes purely radial. 
At this point there is a jump of the logarithm in $\bar P_{\mathrm{eff}}$.  
We draw a schematic picture of the wall trajectory when there is a 
sign flip in Fig.~\ref{fig1}. In this picture the center corresponds
to $R=2GM$, the radial direction is the rescaled radius and the angular 
direction is the Euclidean time, $\bar\tS$. 

\begin{figure}
\begin{center}
\fbox{\includegraphics[width=8cm]{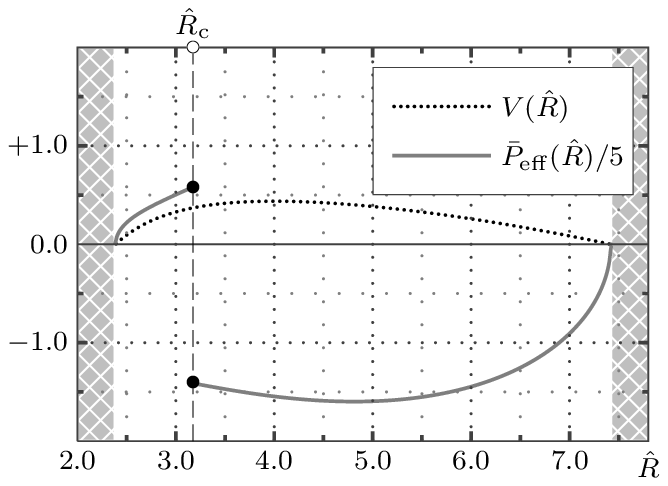}}
\caption{\label{fig2}Plot of the effective potential and of the effective Euclidean
momentum along a tunnelling trajectory. The quantities are calculated for a Minkowski
Schwarzschild junction in which $M = 1$, $\tilde{\sigma} = 0.25$, which results
in the relevant sign for the outside spacetime to change at
$\hat{R} _{\mathrm{c}} \approx 3.175$. The plot clearly emphasizes the discontinuity
in the expression for the effective momentum ({\protect\ref{eq:effmomEuc}})
due to the change in the $\epsilon _{+}$ sign.}
\end{center}
\end{figure}

As a concrete example, let us consider the case of the pure tension wall
with $m=4\pi \sigma R^2$. 
In this case, 
from Eq.~\eqref{solRdot}, we find that the turning points 
corresponding to $\hat R_{,\tau}=0$ are given by the solutions of
\begin{equation}
f(\hat R):=\tilde\sigma^2 \hat R^3-2\tilde\sigma \hat R^2+\Rg =0~, 
\end{equation}
where we have introduced $\tilde\sigma:=4 \pi G \sigma$. 
It is easy to see that $f(\Rg)\geq 0$ and 
the equality holds for $\tilde\sigma=1/\Rg$. 
At the minimum of $f(\hat R)$ where 
$\hat R=4/(3\tilde\sigma)$, we have 
$f(4/(3\tilde\sigma))=\Rg -32/(27\tilde\sigma)$. 
Therefore, we find that there is a classically forbidden region for 
$\tilde\sigma<32/(27\Rg)$. A wormhole can be produced when
the critical radius, where the discontinuity appears,  
\begin{equation}
 \hat R_{\mathrm{c}}=\left({\Rg \over \tilde\sigma^2}\right)^{1/3}~, 
\end{equation}
is in the classically forbidden region. As mentioned above, 
this critical radius does not result in pathologies in the classically allowed region. 
Therefore, if $\hat R_{\mathrm{c}}>\Rg$, the critical radius is under the 
potential barrier. This means that wormhole production 
is possible when $\tilde\sigma <1/\Rg$.  

\begin{figure}
\begin{center}
\fbox{\includegraphics[width=14cm]{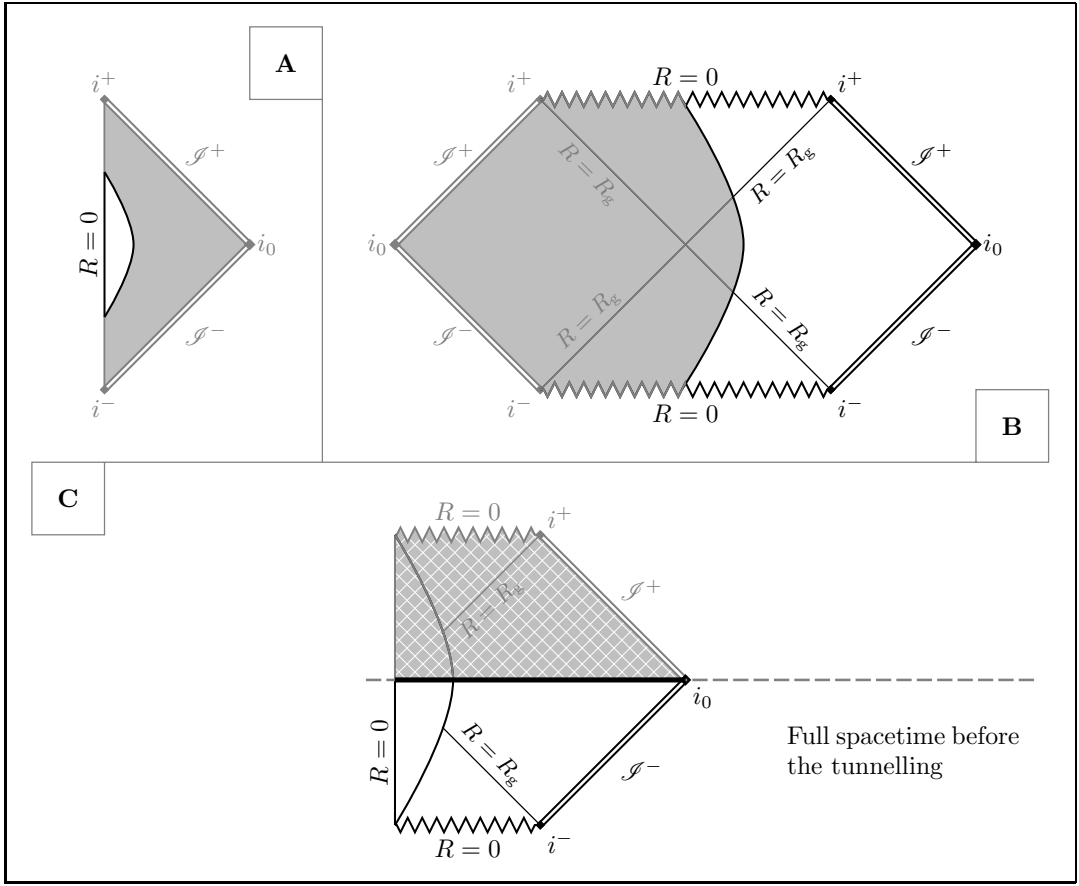}}
\caption{\label{fig3}Construction of the Penrose diagram for the spacetime before
the tunnelling. Panel~A shows the wall trajectory in Minkowski spacetime. The
un-shaded area, between $R = 0$ and the bubble wall, participates in the junction,
and is joined to the un-shaded region of Schwarzschild spacetime in panel~B. The
final configuration is shown in panel~C, where, again, we have to consider only the
un-shaded part of the Penrose diagrams, that describes spacetime while the wall
expands from $R = 0$ until the turning point, where tunnelling takes place. The
thick black line in panel~C is the spacetime slice at which tunnelling starts
(see, \emph{e.g.}, the $\bar{t} _{\mathrm{S}} ^{\,\mathrm{i}}$ slice in
Fig.~{\protect\ref{fig1}}, which corresponds to the Schwarzschild part of this
slice).}
\end{center}
\end{figure}
\begin{figure}
\begin{center}
\fbox{\includegraphics[width=14cm]{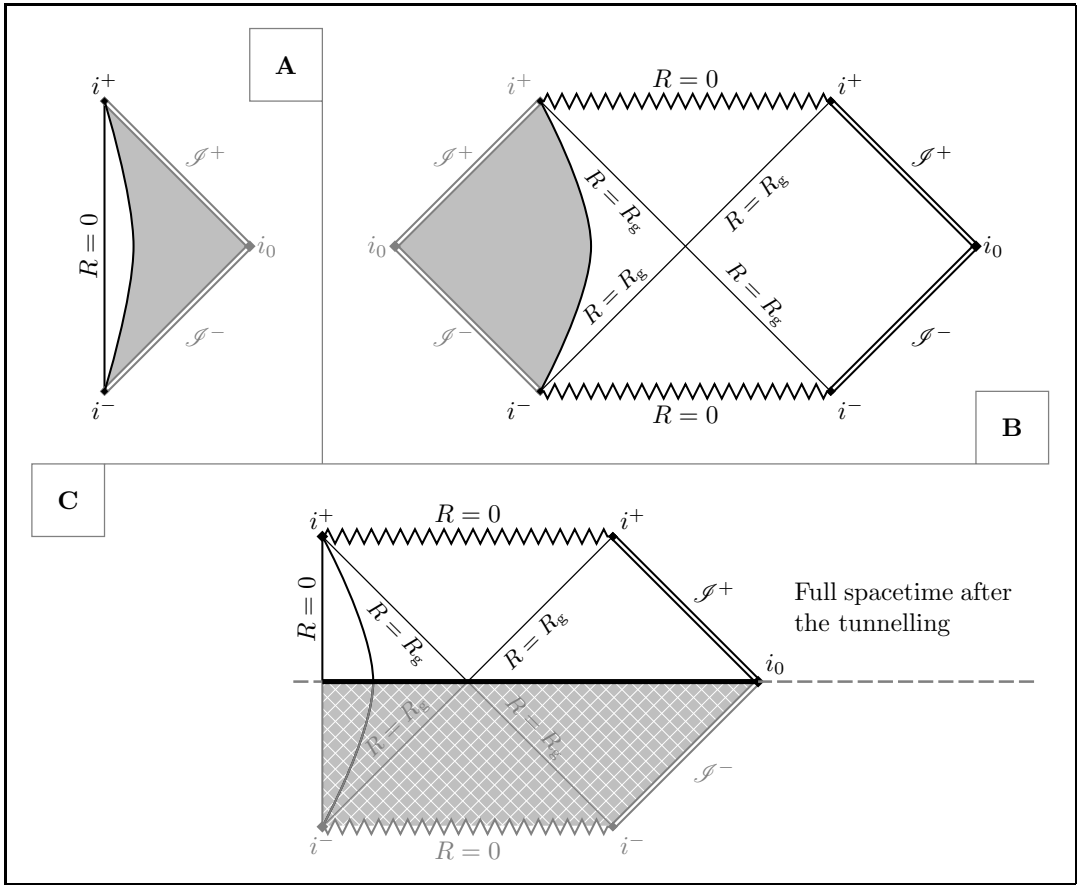}}
\caption{\label{fig4}Construction analogous to the one in Fig.~{\protect\ref{fig3}},
but this time for the spacetime after the tunnelling. The junction is obtained
again by joining the unshaded region of Minkowski spacetime in panel~A, with
the unshaded region of Schwarzschild spacetime in panel~B. After performing the
junction, the spacetime after the tunnelling is the region to the future of the thick
black line in panel~C. The part of this slice in the Schwarzschild region corresponds
to the $\bar{t} _{\mathrm{S}} ^{\,\mathrm{f}}$ slice of Fig.~\ref{fig1}. Here
it is also clear that after the tunnelling the slice contains $R = R _{g}$. This was
not the case for the 
slice before the tunnelling shown in Fig.~{\protect\ref{fig3}}.}
\end{center}
\end{figure}

Now, we discuss the key issue of this paper.  
As long as we use the foliation by the Schwarzschild time, 
{it is problematic to consistently define the Euclidean manifold
interpolating between the configurations before and after the tunnelling.}
As a concrete example, let us consider the case 
shown in Fig.~\ref{fig1} (for this case, plots of the effective momentum along the
tunnelling trajectory and of the potential barrier can  be found in Fig.~\ref{fig2}).  
When $d\bar \tS/d\tau$ is positive, 
the wall is {located at} $\hat{R}<\hat R_{\mathrm{c}}$ and the Schwarzschild
spacetime is relevant for $\hat{R}<R<\infty$. Minkowski spacetime is 
connected beyond the wall. 
After passing through the point $\hat R=\hat R_{\mathrm{c}}$, 
$d\bar \tS/d\tau$ becomes negative. Then,  
the wall is present for $\hat R>\hat R_{\mathrm{c}}$ and the Schwarzschild
spacetime is relevant for $R<\hat R$. Again the Minkowski spacetime is 
connected beyond the wall. Then, one may wonder where
the asymptotic region with $R \to \infty$ is. 
The asymptotic region is on the other side extending beyond 
the center, corresponding to $R=2GM$. The time slice cannot terminate 
at the center (bifurcation point) of the Schwarzschild spacetime. 
We then see that the geometry on this time slice suddenly changes
at the sign flip point. 
Namely, the final configuration contains a wormhole, corresponding 
to the existence of a minimum circumferential radius. 
At the same time $\bar P_{\mathrm{eff}}$ is discontinuous there. 
As long as we stick to this time slice, it is difficult to obtain a 
satisfactory and consistent prescription. Figures~\ref{fig3} and~\ref{fig4}
show the situation before and after the tunnelling, respectively. By comparing
the slice before the tunnelling (thick horizontal line in the Penrose diagram for
the configuration before the tunnelling in Fig.~\ref{fig3}.C), with the slice after the
tunnelling (thick horizontal line in the Penrose diagram for the 
configuration after the tunnelling 
in Fig.~\ref{fig4}.C) we can also have a clear example of the situation
discussed just above for the Euclidean spacetime that should interpolate between
these two configurations.
In the next section we discuss the same process in the canonical formalism 
without specifying the gauge, which makes it
possible to overcome these difficulties. 

\section{\label{sec:CASTS}Canonical approach with smooth time slice}

We consider the canonical approach in this section, following 
Ref.~\cite{Kraus}. 
The spherically symmetric metric is 
specified in the $3+1$ decomposition as 
\begin{equation} 
 ds^2= \Nt dt^2 +L^2(dr+\Nr dt)^2+R^2 d\Omega^2~, 
\end{equation}
where, with standard notation, $d \Omega ^{2}$ is the spherically symmetric part of
the line element.
Then, the action in the canonical formalism is obtained as 
\begin{equation}
  S
  =
  \int dt\, p \, \dot{\hat{r}}
  +    
  \int \! dt \! \! \int dr
      \left({1\over G}
      \{ \pi _{L} \dot{L} + \pi _{R} \dot{R} \}
  - \Nt \Ht -\Nr \Hr \right)
   -  \int d\tS M
  ~,
\label{canonicalaction}
\end{equation}
with 
\begin{eqnarray}
\Ht &=&
     {1\over G}\left(
      {L\pi_L^2\over 2R^2}-{\pi_L \pi_R\over R}+\left({RR'\over
					       2L}\right)'
     -{{R'}^2\over 2L}-{L\over 2}      \right)
     +\delta(r-\hat r)\, \sqrt{(p/\hat
     L)^2+m^2}
          ~,\cr
\Hr &=&
     {1\over G}\left(
     R'\pi_R-L\pi'_L\right)-\delta(r-\hat r)\, p
     ~,
\end{eqnarray}
where $p$, $\pi_L$ and $\pi_R$ are the conjugate momenta to $\hat r$, $L$ and $R$,
respectively. As for derivatives, we adopt the following standard convention:
\begin{equation}
  \dot{B} = \frac{\partial B}{\partial t}
  ~ , \qquad
  B ' = \frac{\partial B}{\partial r}~.
\end{equation}
We stress that 
the values of all the metric functions are assumed to be 
continuous across the wall, although their derivatives can be discontinuous. 
The constraint equations $\Ht=0$ and $\Hr=0$ are solved in the bulk as 
\begin{equation}
  \pi _{L} = R \B
  ~ , \qquad
  \pi _{R} = \frac{\pi _{L} '}{\X}~,
\label{eq:piLpiR}
\end{equation}
where we introduce the following definitions
\begin{equation}
 \X := \frac{R'}{L}~,\qquad 
      \B := ( \X ^{2} - \A ^{2} )^{1/2}~.
\end{equation}
By integrating the constraint equations across the wall, we obtain 
the junction conditions, which in the present notation can be written as 
\begin{equation}
  [ \pi _{L} ] = \frac{Gp}{\hat{L}}
  ~ , \qquad
  [ X ] = \frac{Gm}{\hat{R}} \left( 1 + \frac{p ^{2}}{m ^{2} \hat{L} ^{2}} \right) ^{1/2}
  ~.
\label{eq:juncon}
\end{equation}

In the WKB approximation, the wave function is written as 
$\propto \exp(iI(\hat r,L,R))$ and the conjugate momenta are 
identified as
\begin{equation}
 p={\delta I\over \delta \hat r}~,
\qquad
 \pi_L =G
       {\delta I\over \delta L}~,
\qquad
 \pi_R =G
       {\delta I\over \delta R}~.
\end{equation}
Hence, the action relevant to discuss the WKB wave function is 
\begin{equation}
  I=
  \int dt\, p \, \dot{\hat{r}}
  +{1\over G}\left(
   \int \! dt\!   \int \! dr \, 
      \{ \pi _{L} \dot{L} + \pi _{R} \dot{R} \}\right)
  ~. 
\label{relevantaction}
\end{equation}
We should notice that in this approach removing the last term in 
Eq.~\eqref{canonicalaction} is absolutely unambiguous. 

To handle the above expression \eqref{relevantaction} without specifying the gauge, 
a key observation is the existence of a function $\Phi = \Phi ( L , R , R' )$ that
satisfies 
\begin{equation}
  \delta \Phi \sim \pi _{L} \delta L + \pi _{R} \delta R
  ~, 
\end{equation}
where $\sim$ means that the equality holds neglecting total derivative terms. 
It is then possible to integrate the above equation to obtain
\begin{equation}
  \Phi (L,R,R')
  =
  R R' \log \left(\frac{\X - \B}{\A} \right) 
  +
  R L \B~.
\end{equation}
In the above expression there is an arbitrariness, as a total derivative of an arbitrary
function of $R$ with respect to $r$ can be added, which, of course, does not 
affect the final result. 

Then, the action becomes
\begin{eqnarray}
  I
  & = & \!\!
  \int dt\, p \, \dot{\hat{r}}
  +{1\over G}\left(
  \int \! dr \! \! \int \! dt
    \frac{\partial \Phi}{\partial t}
  -
    \int \! dt
      \left[ \Psi
           \dot{R} \right] \right)
  \nonumber \\
  & = & \!\!
  \int dt\, p \, \dot{\hat{r}}
  +{1\over G}\left(
    \int \! dr\, \Phi\Bigr\vert_{\tSi}^{\tSf} 
  -
  \int \! dt\, \dot{\hat r} [\Phi]
  -
    \int \! dt
      \left[ \Psi \dot{R} \right]\right) ~,
\label{I1} 
\end{eqnarray}
where we define
\begin{equation}
\Psi:=\frac{\partial \Phi}{\partial R'}
=
    R \log \left(\frac{\X - \B}{\A}\right)~. 
\end{equation}
In the first equality of Eq.~\eqref{I1}, we removed the contribution of $\Psi \dot{R}$
at $r \to \infty$, assuming that the time slice is asymptotically 
identical to the one in the static chart of the Schwarzschild spacetime, 
in which $\Psi$ vanishes because $\B=0$ and $\X=\A$. 
Using 
\begin{equation}
  \dot{\hat{R}}
  :=
  \frac{d \hat{R}}{d t}
  =
  \left(R'\dot{\hat{r}}
  +
  \dot{R}\right)_{\!r=\hat r}~,
\label{eq:totder__R}
\end{equation}
we can rewrite the last term in the parentheses
on the right hand side of Eq.~\eqref{I1} as
\begin{eqnarray}
  \left[ \Psi \dot{R} \right]
  & = & \!\!
  \left [
    \Psi \dot{\hat{R}}
    -
    \Psi
    {R'}
    \dot{\hat{r}}
  \right ]
  =   -
  \left [ \Psi {R'} \right ]   \dot{\hat{r}}
  +
  \left [\Psi \right ] \dot{\hat{R}}~, 
\end{eqnarray}
where in the last equality we have extracted out $\dot{\hat{R}}$ and 
$\dot{\hat{r}}$ from the square brackets since their values evaluated 
on both sides of the junction are identical. Thus, we obtain 
\begin{equation}
  I
  =
  \int \! dt \,
    \dot{\hat{r}}
    \left(
      p + {1\over G}
    \left[\Psi {R '} - \Phi \right]
    \right)
    +{1\over G}
    \left(
    \int \! dr\, \Phi\Bigr\vert_{\tSi}^{\tSf} 
  -
  \int \! dt\, \dot{\hat R}\, [\Psi]\right)  
    ~ .
\label{action2}
\end{equation}
As we have 
$\Psi R'-\Phi = - R L \B$,
the first term in Eq.~\eqref{action2} vanishes 
using the junction condition \eqref{eq:juncon}, and
we finally obtain the \emph{gauge unfixed}
action relevant for the WKB wave function as
\begin{equation}
  I
  = {1\over G}\left(
   \int \! dr\, \Phi\Bigr\vert_{\tSi}^{\tSf} 
  -
  \int \! dt\, \dot{\hat R}\, [\Psi]\right) 
  ~.
\label{action3}
\end{equation}

Let us now examine the motion of the shell, $d\hat R/dt$, in more
detail. 
The part of the action related to the shell takes the form
\begin{equation}
  S _{\mathrm{s}}
  =
  \int dt\, 
    L _{\mathrm{s}}
  =
  -m
  \int dt
    \left(
      (\hat \Nt)^{2} - \hat{L} ^{2} ( \dot{\hat{r}} + \hat\Nr ) ^{2}
    \right) ^{1/2}
    ~.
\end{equation}
From this expression, the conjugate momentum to $\hat{r}$ turns out to be
given by 
\begin{eqnarray}
  p
  & = &
  \frac{\partial L _{\mathrm{s}}}{\partial \dot{\hat{r}}}
  =
  m
  \left(
    (\hat\Nt)^{2} - \hat{L} ^{2} ( \dot{\hat{r}} + \hat\Nr ) ^{2}
  \right) ^{-1/2}
  \hat{L} ^{2} ( \dot{\hat{r}} + \hat\Nr )~,
\end{eqnarray}
from which we get 
\begin{equation}
  \frac{\hat{L} ^{2}}{(\hat\Nt)^{2}} ( \dot{\hat{r}} + \hat\Nr ) ^{2}
  =
  \frac{p ^{2}}{m ^{2} \hat{L} ^{2}}
  \left(
    1 + \frac{p ^{2}}{m ^{2} \hat{L} ^{2}}
  \right) ^{-1}
  ~.
\label{rdotN}
\end{equation}
From the normalization of the four velocity, we also find 
\begin{equation}
  \left( \frac{\hat \Nt d\hat t}{d \tau} \right) ^{2}
  \left(
    1 - \frac{\hat{L} ^{2}}{(\hat\Nt)^{2}} ( \dot{\hat{r}} + \hat\Nr ) ^{2}
  \right)
  =
  1~, 
\end{equation}
which is further simplified using Eq.~\eqref{rdotN} as 
\begin{equation}
  \frac{\hat \Nt d\hat t}{d \tau}
  =
  \left(
    1
    +
    \frac{p ^{2}}{m ^{2} \hat{L} ^{2}}
  \right) ^{1/2}
  ~.
\label{eq:fouvelnorcon}
\end{equation}

Now, we are ready to rewrite ${d \hat{R}}/{d \tau}$. 
Using the equation of motion for $R$, 
\[
  \dot{R}
  = 
  - \Nt \frac{\pi _{L}}{R} + \Nr R '~.
\]
Then we obtain 
\begin{eqnarray}
  \frac{d \hat{R}}{d \tau}
  & = & \!\!
  \frac{d \hat t}{d \tau}
  \left(
    ( \hat\Nr + \dot{\hat{r}} ) \hat{R} ' - \hat \Nt \frac{\hat{\pi _{L}}}{\hat{R}}
  \right)
  \nonumber \\
  & = & \!\!
  \frac{\hat\Nt d \hat t}{d \tau}
  \left(
    ( \hat\Nr + \dot{\hat{r}} ) \frac{\hat{L}}{\hat \Nt} \hat{\X} - \hat{\B}
  \right)
  \nonumber \\
  & = & \!\!
  \left( 1 + \frac{p ^{2}}{m ^{2} \hat{L} ^{2}} \right) ^{1/2}
  \left(
    \left( 1 + \frac{p ^{2}}{m ^{2} \hat{L} ^{2}} \right) ^{-1/2} \frac{p \hat{X}}{m \hat{L}}
    -
    \hat{\B}
  \right)
  \nonumber \\
  & = & \!\!
  \frac{p \hat{X}}{m \hat{L}}
  -
  \hat{\B} \left( 1 + \frac{p ^{2}}{m ^{2} \hat{L} ^{2}} \right) ^{1/2}
  ~,
  \label{dhatRdtau}
\end{eqnarray}
where in the third equality, we have used Eqs.~\eqref{rdotN} and \eqref{eq:fouvelnorcon}. 
Substituting $\hat\B= (\hat\X ^{2} -\hat\A ^{2})^{1/2}$, 
this equation can be solved for $\hat \X$ as
\begin{eqnarray}
  \hat{\X}
 =  -
  \frac{p}{m \hat{L}} \hat{R} _{, \tau}
  +\epsilon
  \left(
    \hat{R} _{, \tau} ^{2} + \hat{\A} ^{2}
  \right) ^{1/2}
  \left(
    1 + \frac{p ^{2}}{m ^{2} \hat{L} ^{2}}
  \right) ^{1/2}
  ~.
\label{Xeq}
\end{eqnarray}
Remembering that $p$ and $\hat R_{,\tau}$ do not have
a jump across the junction, 
from Eq.~\eqref{Xeq} and the junction condition \eqref{eq:juncon}, 
we recover exactly Eq.~\eqref{junctionc}. 

Furthermore, substituting Eq.~\eqref{Xeq} into Eq.~\eqref{dhatRdtau}, we obtain
\begin{eqnarray}
  \hat{\B}
  & = &
  -
  \left(
    1 + \frac{p ^{2}}{m ^{2} \hat{L} ^{2}}
  \right) ^{\!\! 1/2}\!\! 
  \hat R _{, \tau}
  +\epsilon
  \frac{p}{m \hat{L}}
  \left(
    \hat{R} _{, \tau} ^{2} + \hat{\A} ^{2}
  \right) ^{\! 1/2}
  ~,
\label{betahat}
\end{eqnarray}
and hence 
\begin{equation}
  \hat{\X} - \hat{\B}
  =
  \left\{
        \left(
      1 + \frac{p ^{2}}{m ^{2} \hat{L} ^{2}}
    \right) ^{\!\! 1/2}\!\!\!\!
   -\frac{p}{m \hat{L}} \right\}
  \left\{
    \epsilon
    \left(
      \hat{R} _{, \tau} ^{2} + \hat{\A} ^{2}
    \right) ^{\! 1/2}\!\!\!
    +
    \hat{R} _{, \tau}
  \right\}
  ~.
\label{Xminusbeta}
\end{equation}
Therefore, we can finally write the jump of $\Psi$ as
\begin{equation}
[\Psi]=\hat R \left[\log\left({ \epsilon \left(
      \hat{R} _{, \tau} ^{2} + \A^{2}
    \right) ^{\! 1/2}\!\!\!  + \hat{R} _{, \tau}\over 
    \A
}\right)\right]~. 
\end{equation}
After Euclideanization, Eq.~\eqref{action3} can be then rewritten
using the above results, and it gives 
\begin{equation}
  \bar I
  = {1\over G}\left(
   \int \! dr\, \bar\Phi\Bigr\vert_{\bar\tSi}^{\bar\tSf} 
  -
  \int \! dt\, \dot{\hat R}\, [\bar\Psi]\right) 
  ~, 
\label{IE}
\end{equation}
with 
\begin{eqnarray}
  \bar\Phi
  =i\Phi
  =
  i R R' \log \left(\frac{\X -i\left(
{\A^2-\X^2}\right)^{1/2}}{\A} \right) 
  -
  R L \left({\A^2-\X^2}\right)^{1/2}~,
\end{eqnarray}
and 
\begin{eqnarray}
[\bar\Psi]=
[i \Psi]=
i\hat R \left[\log\left({ \epsilon \left(
      \A^{2}-\hat{R} _{, \bar\tau} ^{2} 
    \right) ^{\! 1/2}\!\!\!  + i \hat{R} _{, \tau}\over 
    \A
}\right)\right]~. 
\end{eqnarray}

This expression 
is identical to Eq.~\eqref{Iconventional} 
obtained in Sec.~\ref{sec:conventional} for 
the tunnelling that does not produce a wormhole.  
First, since $\A=\X$ on the initial and 
final surfaces, where the time slices coincide 
with the ones with $\bar \tS=$constant and $\bar \tM=$constant, 
$\bar\Phi$ vanishes there. 
Since $\epsilon=+1$ in this case, as mentioned earlier, 
the difference between $\bar P_{\mathrm{eff}}$ and $[\bar\Psi]$ 
does not arise. 

By contrast, in the case with wormhole production 
the first term in Eq.~\eqref{IE} does not vanish because 
$\X$ is negative in the region between $R=\Rg$ and the wall in Schwarzschild,
and hence $\X=-\A$ there. Namely, the first term contributes as 
\begin{equation}
\int \! dr\, \bar\Phi\Bigr\vert_{\bar\tSi}^{\bar\tSf} 
 =\int_{\hat r(\bar\tSf)}^{\rg} \! dr\, \pi RR'
 ={1\over 2}\left(\Rg^2-{\hat R(\bar\tau^{\mathrm{f}})}^2\right)~, 
\end{equation}
where $\rg$ is the value of $r$ at $R=\Rg$ on the final surface. 
Hence, the difference between Eq.~\eqref{Iconventional} 
and \eqref{IE} is evaluated as
\begin{equation}
\bar I-\bar I_{(\tS)}=\int \! dr\,
		 \bar\Phi\Bigr\vert_{\bar\tSi}^{\bar\tSf} 
         +\pi \int_{R_c}^{\hat R(\bar\tSi)} \! dR R
   ={1\over 2}\left(\Rg^2-R_c^2\right)~, 
\end{equation}
if we assume that $\bar P_{\mathrm{eff}}$ in Eq.~\eqref{Iconventional} 
has a discrete jump at $\hat R=R_{\mathrm{c}}$. Of course, this discrepancy 
is not strange at all, since the naive 
extension of the validity range of the formula \eqref{Iconventional} 
cannot be justified. 

\section{\label{sec:CDE}Consistent direct evaluation}

As we anticipated, we will now show that the method using a 
\emph{pseudo-manifold} for the description of the instanton solution 
gives the same result that we derived by using the canonical approach 
in the preceding section. Although this equivalence might seem almost trivial because 
both approaches are based on the same smooth foliation of Euclidean spacetime,
its explicit proof would be pedagogically useful. 

We return then to the discussion in Sec.~\ref{sec:conventional}. 
The first key observation is that the contribution from 
the carefully included 
$I _{\mathrm{surface}} ^{\mathrm{wall}}$ 
should not be included when we adopt a smooth foliation. 
The second point is that we have rewritten a term in 
Eq.~\eqref{gravityterm} as
\begin{equation}
   \int _{\tau ^{\mathrm{i}}} ^{\tau ^{\mathrm{f}}} d \tau
      \frac{\hat{R}^2(\hat\AS)^2_{,\hat R}}{4G\BETA}
   =   -\frac{M}{2}  
       \int _{\tau ^{\mathrm{i}}} ^{\tau ^{\mathrm{f}}} d \tau
      {d\hat{\tS}\over d\tau}~. 
\label{lastterm}
\end{equation}
We have then subtracted $M(\tSf-\tSi)$ from the total action. 
In the computation of Sec.~\ref{sec:conventional} 
half of this subtraction was compensated by $I _{\mathrm{surface}} ^{R _{\mathrm{BIG}}}$
and the rest by the above contribution~\eqref{lastterm}. 
However, we find 
\begin{equation}
\int _{\bar\tau ^{\mathrm{i}}} ^{\bar\tau ^{\mathrm{f}}} d \bar\tau
      {d\hat \tS\over d\bar\tau}
      =\bar\tSf-\bar\tSi+2\pi \Rg~,
\end{equation}
when we use a smooth foliation for the tunnelling solution with wormhole
formation. This shows that an additional contribution $\pi M \Rg$
to the Euclidean action arises. Gathering all, we find that 
the Euclidean action evaluated by using a smooth foliation is given by  
\begin{eqnarray}
\bar I_{(\tS)}-\bar I _{\mathrm{surface}} ^{\mathrm{wall}}  +\pi M \Rg 
& =&{1\over G} \left(
    - \int \! dt\, \dot{\hat R}\, [\bar\Psi]
   +
    {\hat R\over 2} \hat{\bar\Phi}\Bigr\vert_{\bar\tSi}^{\bar\tSf} 
   +{\Rg^2\over 2}\right) 
\cr
& =&
{1\over G} \left(
    - \int \! dt\, \dot{\hat R}\, [\bar\Psi]
   +{\Rg^2-\hat R(\bar\tau^{\mathrm f})\over 2}\right)~,  
\end{eqnarray}
which is precisely identical to $\bar I$.

\section{\label{sec:sumcon}Summary and Discussion}

In this work we studied the wormhole production for the simplest spherically 
symmetric shell model in asymptotically flat spacetime. 
In this simple setup, the instanton solution can be generically described by 
the junction of Euclideanized Minkowski and Schwarzschild spacetimes. 
This solution, however, is not a Riemannian manifold in the sense
that the existence of the domain wall may depend on 
the path taken to reach the possible location of the wall in spacetime.
The term \emph{pseudo-manifold} was used
in~\cite{Farhi} to denote this solution. A key point that we have emphasized
here, is that in this case the ordinary constant-time
surfaces associated with the static chart do not foliate the 
instanton smoothly. As a result, methods based on this time slicing 
inevitably become conceptually ambiguous.

We have here discussed, however, that even in these cases, if we choose a
smooth time slicing to connect the configurations before and after the 
tunnelling, it is still possible to find the WKB wave function along an
interpolating path of configurations with a bubble wall. In this way, we can 
identify an appropriate expression for the tunnelling rate without any 
ambiguity. The result agrees with the direct evaluation of the 
Euclidean action once we properly subtract the zero-point 
energy and count how many times each region in the instanton 
solution is swept when we consider a smooth foliation. 

It is possible to trace the subtle nature of the \emph{pseudo-manifold} to
the fact that the time lapse in the Euclidean region is not positive everywhere.
Indeed, the sign of the time lapse has to be opposite between the center and the
asymptotic infinity, for at least some range during the time evolution. This is a
feature that is common to the upward tunnelling in the case of bubble nucleation. 
It would be worth investigating whether or not this negative lapse 
causes any problem when we take into account fluctuations around the WKB
trajectory.

\acknowledgments 

This work was supported in part by the Grant-in-Aid for Scientific Research
(Nos. 24103006, 24103001 and 26287044). One of us, S.~A., would like to heartfully
thank the Department of Physics of Kyoto University, for extended hospitality and
support.

\end{document}